\documentclass[12pt,a4paper]{article}

\usepackage{amsmath}
\usepackage{accents}
\usepackage[T1,T2A]{fontenc}
\usepackage[cp1250,cp1251]{inputenc}
\usepackage[english]{babel}
\usepackage{graphicx}
\usepackage{amsfonts}
\usepackage{amssymb}
\usepackage{stmaryrd}
\usepackage[T1]{fontenc}
\usepackage{calligra}
\usepackage[toc, page]{appendix}
\usepackage{flafter}
\usepackage{wrapfig}

.pk scaled1212
.pk scaled1212
.pk scaled1212

\allowdisplaybreaks[1]

\inputencoding{cp1251}

\newcommand\ntrl{{\mathbb N}}
\newcommand\intg{{\mathbb Z}}

\newcommand\real{{\mathbb R}}
\newcommand\cplx{{\mathbb C}}

\newcommand{\ben}{\begin{equation}}
\newcommand{\bal}{\begin{align}}
\newcommand{\een}{\end{equation}}
\newcommand{\eal}{\end{align}}
\newcommand{\bea}{\begin{eqnarray}}
\newcommand{\eea}{\end{eqnarray}}

\newcommand{\qq}{\qquad\qquad}
\newcommand{\QQ}{\qquad\qquad\qquad\qquad}

\newcommand{\hfb}{{\hfill\break}}

\newcommand{{\ga}}{{\gamma}}

\input amssym.def
   \def\N{{\Bbb N}}

\begin{document}

\parskip=4pt
\baselineskip=14pt


\title{\vskip-1cm Flow of  $S$-matrix poles for elementary quantum
potentials\footnote{This research was supported in part by an NSERC Undergraduate Summer Research Award (SN) and an NSERC
Discovery Grant (MW).}}
\author{B. Belchev, S.G. Neale, M.A. Walton \\\\ {\it Department of Physics and Astronomy,
University of Lethbridge}\\
{\it Lethbridge, Alberta, Canada\ \  T1K 3M4}\\\\
{\small borislav.belchev@uleth.ca, samuel.neale@uleth.ca, walton@uleth.ca}\\\\
}

\maketitle

\begin{abstract}
The poles  of the quantum scattering matrix ($S$-matrix) in the complex momentum plane  have been studied extensively. Bound
states give rise to $S$-matrix poles, and other poles correspond to non-normalizable anti-bound, resonance and anti-resonance
states. They describe important physics, but their locations can be difficult to find. In pioneering work, Nussenzveig
performed the analysis for a square well/wall, and plotted the flow of the poles as the potential depth/height varied. More
than fifty years later, however, little has been done in the way of direct generalization of those results. We point out that
today we can find such poles easily and efficiently, using numerical techniques and widely available software. We study the
poles of the scattering matrix for the simplest piece-wise flat potentials, with one and two adjacent (non-zero) pieces. For
the finite well/wall  the flow of the poles as a function of the depth/height  recovers the  results of Nussenzveig. We then
analyze the flow for a  potential with two independent parts that can be attractive or repulsive, the two-piece potential.
These examples provide some insight into the complicated behavior of the resonance, anti-resonance and anti-bound poles.
\end{abstract}


\vfill\eject

\newpage


\section{Introduction}

The scattering matrix has a rich, interesting history. It has even, in the past, been  postulated to provide a fundamental
physical viewpoint \cite{history}. During the first half of the 20th century quantum field theory was `plagued' by infinities.
Many were skeptical of the prospects of quantum field theory to explain  physical reality. As a result the $S$-matrix became of
central importance and was extensively studied. Once renormalization techniques resolved the problems with the infinities and
more and more elementary particle phenomena were successfully explained using quantum field theory, the $S$-matrix lost its
fundamental role.  It  remains important in high energy physics and other fields, however, and is still an interesting object
of study. For one, it provides a useful perspective on the quantum physics of local interactions. At a more fundamental level,
the poles of the $S$-matrix  provide a unified description of stable and decaying states.  Also, they are actively studied by
mathematicians in relation to various aspects of spectral theory \cite{Simon_Reed}.

It is well known that bound states correspond to poles of the scattering matrix. However, the $S$-matrix has other poles that
are not associated with normalizable states, but still encode important and interesting physics \cite{resonances,BZP}. In this
paper we consider the positions of all the poles of the $S$-matrix for certain elementary potentials, and their flow under
deformation of the potential parameters.

What is the physical significance of the $S$-matrix poles that are not associated with bound states? In the late 1920's Gamow
proposed an explanation of $\alpha$-decay in terms of solutions of Schr\"odinger's stationary equation with complex
eigenvalues, that satisfied a purely outgoing boundary condition, i.e. far enough from the origin the solutions were outgoing
plane wave. These solutions could be thought of as wave functions, i.e. as the position representations of certain generalized
state vectors. Those same {\it Gamov vectors} turned out to correspond to the poles of the $S$-matrix and the residues of the
propagator. What Gamow sought to apply to $\alpha$-decay was in general a way of describing bound and quasi-stable states that
emphasizes their similarities. The concept of a quasi-stable state is a fundamental one, and so it has been applied
extensively, and in all areas of physics.\footnote{Of course, that continues to this day. To mention some recent applications,
they have been used to calculate tunneling ionization rates \cite{ion}, to understand the phenomenon of diffraction in time
\cite{DIT}, to describe the decay of cold atoms in (quasi-)one-dimensional traps \cite{dCM}, and are directly relevant to
recent condensed-matter experiments \cite{CMPres}. }

The physical effects of the $S$-matrix poles unrelated to bound states are undeniable in scattering (see \cite{GSS}, e.g.).
Strictly speaking, however, the quasi-stable states associated with those poles, the resonance and anti-bound states, are not
true states.\footnote{For brevity and simplicity, however, we will continue to refer to them as states, rather than as `states'
or virtual states, and rely on the context to make the distinction. See \cite{OG} for a discussion.}  The non-normalizability
of their wave functions is one marked difference from the physical bound states. This non-unifying characteristic, however, can
be tamed somewhat, in a mathematical way, by regularizing the integrations,\footnote{This was first done by Zel'dovich; see
\cite{BZP}.} interpreting the probability in a time-dependent setting \cite{Hatano}, and/or continuing to complex potentials
(see \cite{ECS}, e.g.). Upon continuation to complex potentials, one can also relate the different poles to each other (bound
state poles to anti-bound state poles, e.g.) \cite{Kawasaki}.

We should note that, in spite of their non-normalizability, the wave functions associated with resonance and anti-bound states
can be useful. After Gamow's work, they figured in expansions of the $S$-matrix (\cite{Siegert}, e.g.) and propagators
(\cite{Peierls}, e.g.) and other physical quantities. As a result, various ways of determining and approximating them have been
developed (see \cite{WF}, e.g.). We will not consider them here, however.

We focus on the $S$-matrix. Remarkably, it can be determined by its poles--their locations and residues \cite{resonances}. The
pole locations, and their flow on deformation of the potential, are therefore clearly of interest. In the simplest case of a
square well or wall (barrier), the movements of the poles in the complex momentum plane were studied in
\cite{Nussenzveig}.\footnote{In fact Nussenzveig did more--a spherically symmetric rectangular potential well/wall for zero
angular momentum was worked out. Some results were also given for higher angular momenta.}   The depth of a simple square well
and the height of a square barrier were varied to change the position of the $S$-matrix poles and generate their flow. More
recently, certain of Nussenzveig's results were reproduced using simple, elegant graphical methods in \cite{Moiseyev}, and
different poles were related by complex continuation in \cite{Kawasaki}.

More than fifty years after Nussenzveig's work, however, little has been done in the way of direct generalization. Even for the
next-simplest  potentials, numerical methods must be used to find the $S$-matrix poles.  But today, it is not difficult to
perform the analysis in \cite{Nussenzveig} on a personal computer, using widely-available software (such as Mathematica, Maple,
MATLAB, e.g.). To illustrate this point, we will reproduce those results here and generalize them to the next-simplest cases.
We will plot the trajectories of all poles for two types of elementary piece-wise flat potentials--attractive/repulsive
(following \cite{Nussenzveig}) and a two-piece combination of both. For short, we will refer to the potentials as one-piece, or
wall or well or wall/well,  and two-piece, or well+wall, etc.

The above potentials are interesting because they are simple, but still generic. Their simplicity permits a relatively easy
numerical treatment of the pole structure and a complete description of the pole positions in the complex momentum plane, with
a  flow that is  a function of  only a few parameters. Most importantly, unlike certain special potentials such as the Dirac
delta and the Coulomb potential, they are generic enough to exhibit all the possible families of poles.

We should mention that another way to generalize the Nussenzveig flow is to study potentials for which the $S$-matrix can be
found exactly (see \cite{Rawit}, e.g.).   Such solvable potentials are special, however, and so may show non-generic
properties. Here, we prefer to generalize the wall/well in a very direct, simple way that we hope will not introduce any
special features into the flow.


\section{Background}
\begin{figure}
\begin{center}
 \includegraphics[width=100mm]{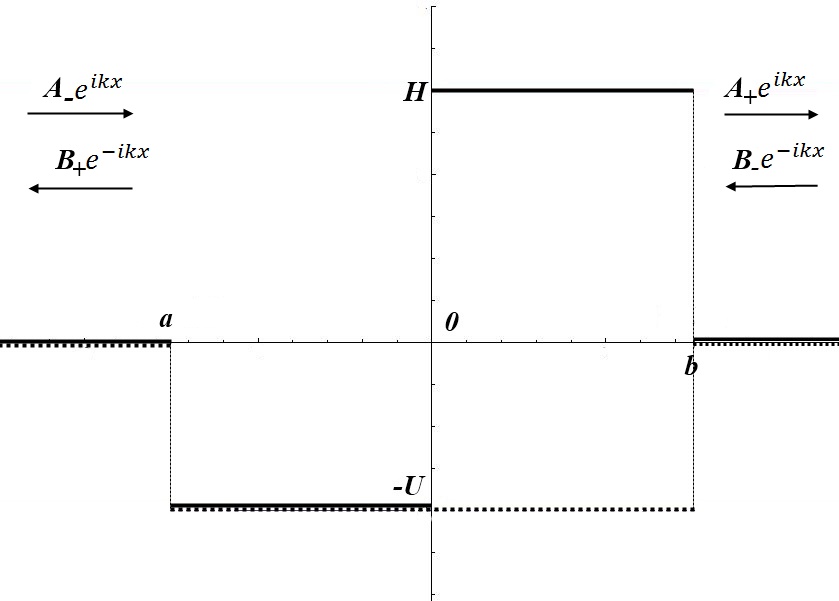}
\caption{\small\textit{\bfseries Potentials studied. }
\it{The dashed line shows the square well/wall  potential with variable depth/height $U>0$/$U<0$.
The solid line represents a 2-piece flat potential with an attractive part and repulsive part, the well+wall.} }
\label{fig:potentials}
\end{center}
\end{figure}

\begin{figure}
 \begin{center}\includegraphics[width=140mm]{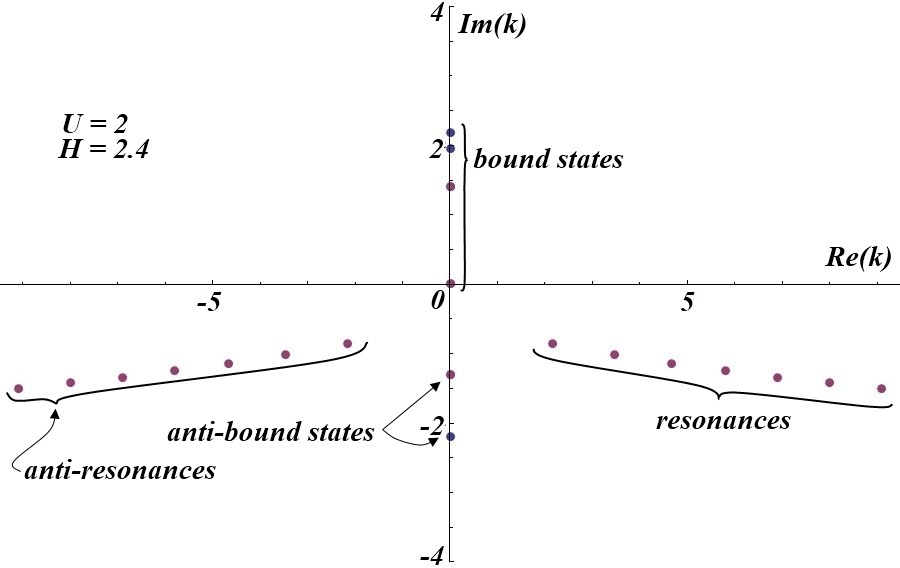}
\caption{\small\textit{\bfseries Types of $S$-matrix poles in a typical configuration. }
\it{Different types of poles correspond to wave functions with different boundary conditions.}}
\label{fig:poles}
\end{center}
\end{figure}
Let a spin-zero, massive particle move on the real line with coordinate $x$. The Hamiltonian of the system  $H=p^2/2m +V(x)$
has a potential $V(x)$ that vanishes fast enough\footnote{We will adopt the same convention as \cite{BZP}, ch. 3, i.e.  faster
than $1/x$. That guarantees asymptotic solutions of the type $e^{\pm i k x}$.} at infinity for scattering to be possible. The
stationary Schr\"odinger equation \ben \left[-\hbar^2\partial_x^2/2m + V(x)\right]\psi=E\psi.\label{eigenvalue}\een determines
the wave functions of the stationary states and their energies. Let us define the momentum variable (wave number) $k$ so that
$k^2= {2mE}/\hbar^2$, as it will have an important role in this work. We will set $m= \hbar=1$ in anticipation of the numerical
results and to lighten the notation. Anytime we consider scattering we deal with plane waves, whether as states or as boundary
conditions, which makes it more convenient to work with the  momentum $k$ instead of the energy $E$.

Some of the  poles of the $S$-matrix can be identified with the bound states of the system: their locations in the $k$-plane
correspond to the bound state energies. The remaining poles turn out to be physically significant, as well. They correspond to
the so-called {\it anti-bound} and {\it resonance states}, and half\footnote{Resonances occur in pairs, at momenta $\pm
k_1-ik_2$, where $k_1,k_2>0$. To distinguish the two poles of such a pair, those in the fourth (third) quadrant are known as
(anti-)resonances.} of the latter are sometimes called {\it anti-resonance states}. The poles will be the central objects of
interest of this paper. We therefore provide a short summary of relevant results.

Let us consider scattering by a potential that vanishes outside of a finite interval\footnote{Then it clearly decreases `fast
enough' at infinity.} $[a,b]$, where $a<0,b>0$. The asymptotic wave function then has the form \bea\label{asymptotic_form} \psi
(x)\rightarrow A_\pm e^{ i k x}+B_{\mp}e^{ - i k x}, \ x\rightarrow\pm\infty,\ x\notin[a,b]. \eea The map that relates the
incoming $(-)$ and outgoing $(+)$ coefficients is the $S$-matrix: \bea\label{S_defined} \left(\begin{matrix} A_+\cr B_+\cr
\end{matrix}\right)=S \left(\begin{matrix} A_-\cr B_-\cr
\end{matrix}\right)
\eea Now, let us define the Jost functions $\phi_\pm$,  as the two independent solutions that behave asymptotically as $e^{\pm
i k x},$ respectively, at $\pm\infty$. The nonzero Wronskian  shows that $\phi_\pm(x,k)^*=\phi_\pm(x,-k)$ is independent of
$\phi_\pm(x,k)$ and therefore we can express $\phi_+$ via $\phi_-$ and its complex conjugate, and vice versa: \ben \phi_+(x,
k)=\alpha(k)\phi_-(x,-k)+\beta(k) \phi_-(x,k) \label{Jost_via_Jost}. \een One can write a similar equation to express $\phi_-$
with different coefficients. Evaluating the Wronskians of each side with the appropriate choice of  $\phi_\pm(x,\pm k)$,
reveals that those coefficients can be expressed using $\alpha(k)$ and $\beta(k)$ \cite{resonances}.

The components of $S$ are related to the well  known {\it transmission } and {\it reflection coefficients (amplitudes)} that
are usually defined via the wave functions for a wave incident from the left or right: \bea\label{reflection_WF} \psi_L(x)=
\Biggl\lbrace
\begin{matrix} e^{ i k x}+R_{+}e^{ - i k x}  { \,\,\, \rm for} \  x\ll a\ , \cr
T_+e^{ i k x}\ \  \ \  \ \  \ \  \ \   \ \ {\rm for} \  x\gg b\ ;\cr
\end{matrix} \ \  \
\psi_R(x)=\Biggl\lbrace
\begin{matrix} T_-e^{- i k x}  \ \  \ \  \ \  \ \  \ \ { \ \ \rm for} \  x\ll a\ , \cr
e^{- i k x}+R_{-}e^{ i k x} \ \ {\,\rm for} \  x\gg b\ .\cr
\end{matrix}
\eea Let us then consider a wave incident from the left. Notice that  $\psi_L(x,k) =\phi_-(x,-k)+R_+(k) \phi_-(x,k)$ and
$\psi_L(x,k)  =T_+(k)\phi_+(x,k)$ satisfy the Schr\"odinger equation. Moreover, expressing $\phi_+$  in terms of  $\phi_-$
using (\ref{Jost_via_Jost}) shows \ben T_+=1/\alpha(k),\ \ R_+(k)=\beta(k)/\alpha(k). \label{transmission_reflection} \een
Using   $\psi_R$ as the wave function and proceeding just as we did with $\psi_L$ reveals that the transmission coefficient
$T_-$ coincides with $T_+$, so we define $T(k):=T_-(k)=T_+(k)$. Also, in agreement with the conservation of probability, $R_-$
only differs from $R_+$ by a phase and can be found, performing a similar calculation for a wave incident from the right. Using
$\psi_L$ and  $\psi_R$ as the wave function in (\ref{asymptotic_form}) and writing the corresponding values of the coefficients
$A_\pm$ and $B_\pm$, via  (\ref{S_defined}), for each case allows us to express the  $S$-matrix in the form
\begin{align}\label{S_matrix}
S=\left(\begin{matrix}
T &R_+\\
R_-& T
\end{matrix}\right),
\end{align}
 a  more familiar form, convenient for practical calculations.

One can show (e.g. \cite{Faddeev})  that the Jost functions are analytic functions of $k$ in the upper-half $k$-plane. The
functions  $\alpha(k)$ and $\beta(k)$, therefore, are also holomorphic for Im$(k)>0$. It is clear now that the poles of the
$S$-matrix are fully determined by the zeros of  $\alpha(k)$.

Now let us assume that $\alpha(k_0)$=0, for $k_0$ somewhere in the upper plane. From  (\ref{Jost_via_Jost}) it follows that for
$k=k_0$ the Jost functions are linearly dependent. Now, recall that the Jost functions are solutions of the Schr\"odinger
equation and  since they are linearly dependent for $k_0$ we have a solution that vanishes  asymptotically at both infinities,
i.e. a normalizable, bound state exists with energy $k_0^2$. However, the Hamiltonian $H$ is assumed to be self-adjoint, and
therefore has only real eigenvalues. Since $\vert \alpha(k)\vert^2=1+\vert\beta(k)\vert^2$, $\alpha(k)\ne0$ for any $x\in\real$
and thus the energy can only be real for $k_0$ purely imaginary.

In addition $\alpha^*(k)=\alpha(-k)$ (from $\phi_+(x,k)^*=\phi_+(x,-k)$) and the Schwarz reflection principle\footnote{The
Schwarz reflection principle states that for a holomorphic function  in the upper complex plane, continuous and real valued on
the real line, one can write an analytic continuation for the whole $\cplx$ plane such that  $f(z^*) = f^*(z)$.} imply that the
analytic continuation of $\alpha(k)$ to the lower part of the complex plane  is symmetric across the Im$(k)$-axis.

The complexification of the momentum,  and therefore the energy, is more than a  mathematical ``trick'', however.  It has sound
physical meaning. In the introduction we mentioned wave functions corresponding to complex eigenvalues and the corresponding
Gamow vectors. The real parts of the complex eigenvalues $z=E -i \Gamma/2$ correspond to the physical energies, and their
imaginary parts correspond  to the decay widths, so that $1/\Gamma$ is the lifetime of the decaying state \cite{Madrid_II}. The
Gamow vectors  satisfy  Schr\"odinger's equation with outgoing boundary conditions (see, e.g. \cite{resonances},
\cite{Madrid_I} and \cite{Madrid_II}).

Our analysis so far does not imply anything about the existence of poles in the lower complex $k$-plane.  The only requirement
is that the poles have to be symmetric with respect to reflection across the Im$(k)$-axis.  In particular, they can still lie
on it, much like the poles associated with the bound states. If the potential is such that the $S$-matrix has poles in the
lower plane, is there a pole-state correspondence as in the bound state case?

As already mentioned, the states signalled by resonances are the Gamow vectors. This is  easy to see in the simplest case of a
square well. As Zavin and Moiseyev \cite{Moiseyev} show, outgoing boundary conditions lead to the same equations that determine
the poles of the $S$-matrix. In addition,  the anti-bound states can be obtained using incoming wave boundary conditions.

Let us point out that all poles except those on the positive Im$(k)$-axis lead to non-normalizable wave functions. A way of
dealing with the spatial divergence is considering the time dependent wave functions: \bea \psi_z(x,t)=e^{-iEt}e^{-\Gamma
t/2}\psi_z(x),\label{res_time} \eea which obviously decay with time, as long as $\Gamma>0$. In terms of the wave number $k$ we
have $\Gamma = -4$Re$(k)$Im$(k)$>0 only for Re$(k)$>0, i.e. for resonance poles. The anti-resonances, with Re$(k)$<0,
correspond to the time-reversed behaviour of the resonances. The interpretation of the remaining divergent wave functions,
those of the anti-bound states, with Re$(k)$=0, remains obscure, though they are closely related to the bound states and
resonances. In addition, as indicated by Nussenzveig even the resonance poles' interpretation breaks down for $\vert \Gamma/
E\vert\gg1$.

The complex eigenvalues do not contradict the well known reality of the eigenvalues of self-adjoint operators. The Hamiltonian,
like any operator,  is defined not just by its action but also by its domain. Strictly speaking then, two operators with the
same action can have two different domains, which makes them two distinct operators. The domain for which we have determined
the self-adjointness of the Hamiltonian, does not contain  those complex energy states, since they are not normalizable in the
usual sense. If we include the Gamow vectors in the domain the operator will no longer be self-adjoint.

To conclude this preliminary section let us make a connection with another important quantity -- the resolvent. Resonances are
also studied through its poles. The reason is that, while the $S$-matrix does not exist for all Hamiltonians, the resolvent is
a  more general mathematical object and as such can be defined for much wider class of operators. And when scattering is
possible, the poles of the  $S$-matrix coincide with those of the  resolvent.

Consider the time evolution of the state vector as determined by the propagator, the unitary operator of the form
$U(t_1,t_2)\,=\, e^{-i(t_2-t_1)H/\hbar}$ for $t$-independent Hamiltonian $H$. Its Laplace transform defines the {\it resolvent}
operator of the Hamiltonian, \bea R_H(E)=(H-E {\bf 1})^{-1}=\sum_{\alpha\in I}\frac{\vert E_\alpha\rangle\langle E_\alpha
\vert}{E_\alpha-E},\label{resolvent} \eea where $I\subset\real$ is a discrete or continuous interval depending if the energies
are part of the discrete or continuous spectrum, respectively. Comparing (\ref{eigenvalue}) and (\ref{resolvent}) shows that
eigenvalues of the Hamiltonian generate poles of the resolvent.  The resolvent, however, treats all the complex eigenvalues  on
an equal footing.

The resolvent and the $S$-matrix, as a function of $k=\sqrt{2mE}/\hbar$, have the same singularities for all physical purposes,
as we mentioned.  This is to be expected since  $R_H(E)$ is related to the Green's function $G(x,y)$: the Schwarz operator
kernel of $R_H(E)$ is the Green's function $G(x,y)$, satisfying $(H-E)G(x,y)=\delta(x-y)$ (see, e.g. \cite{Faddeev_Seckler}).
For scattering a natural assumption is that the initial state was the wave function defined in the far past, i.e. at
$t=-\infty$, and similarly the final state, at $t=+\infty$. Thus the resolvent is also related to the scattering matrix  via
$S=U(-\infty,+\infty)$ and a Laplace transform.  \footnote{For the mathematical rigour that we ignore, the reader may refer to
the mathematical literature. Quite comprehensible lecture notes are available by Tang and Zworski \cite{Zworski},  for
example.}

For our purposes, however, finding the pole content of the scattering matrix is simpler than finding that of the resolvent. For
piece-wise flat potentials it can be done  by simply matching the values of the wave function and its derivative at the
discontinuities. Simple algebraic equations result, that can be handled easily by software. The resolvent makes the
relationship between its poles and the spectrum clearer but the $S$-matrix allows for easier calculations.


\section{Flows for the finite square well/wall}
\begin{figure}[t]
 \begin{center}
 \includegraphics[width=110mm]{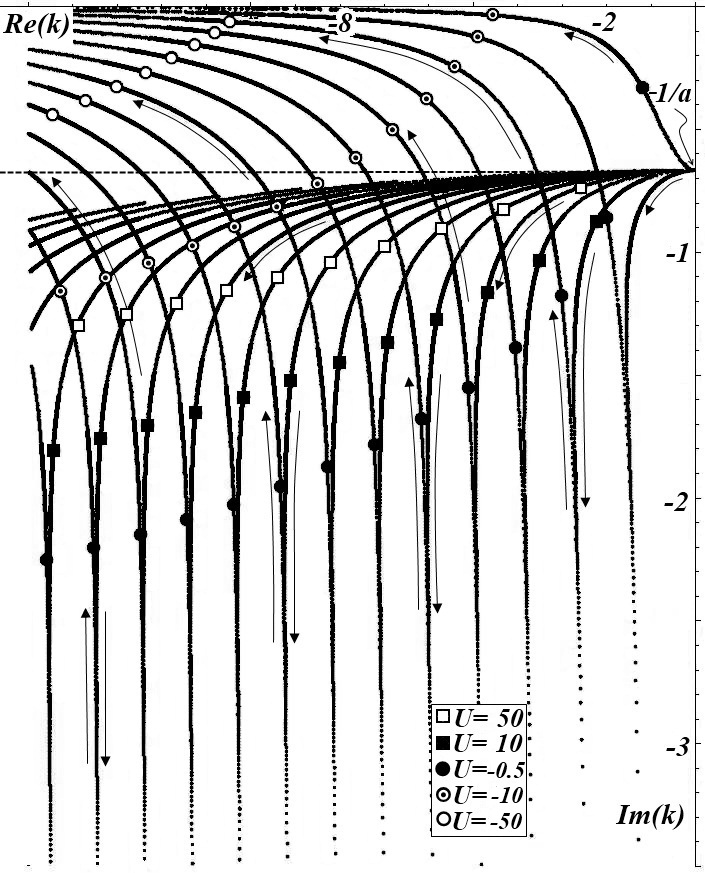}
\caption{\small\textit{\bfseries{Flow trajectories of the anti-resonance poles for the potential well/wall with variable depth/height. }}
\it{Only the  trajectories for ${\rm Re}(k)<0$ are shown since there is a  complete  symmetry under reflection across the
${\rm Im}(k)$-axis. The arrows indicate the direction of the flow for $-U$  varying from $-\infty$ to $\infty$}(a deep well to a tall wall).}
\label{fig:Resonance_symmetric}
 \end{center}
\end{figure}

 A  particle moves freely along the real line except in $[-a,a]$;  it interacts by the piece-wise flat potential
\bea\label{well}
V(x)=\Biggl\lbrace
\begin{matrix} &-U, \ x \in[-a,a],\qq\ \ \ \ \cr
&0,\  x\in(-\infty,-a) \cup (a,\infty). \cr
\end{matrix}
\eea
\begin{figure}
 \begin{center}
 \includegraphics[width=99mm]{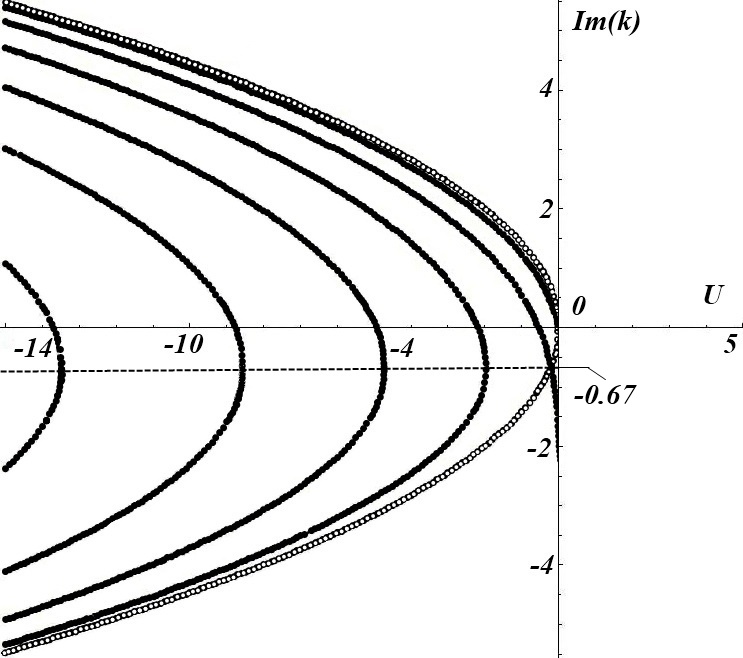}
\caption{\small\textit{\bfseries{Flow trajectories of the  bound  and anti-bound states for a potential well (wall) with depth (height)
-U>0 (U>0). }}\it{The hollow points trace the positions of the branch points. The dashed line indicates the coalescence point $k=-i/a$.}}
 \label{fig:symmetric_bound}
\end{center}
\end{figure}
\begin{figure}
\begin{center}
 \includegraphics[width=120mm]{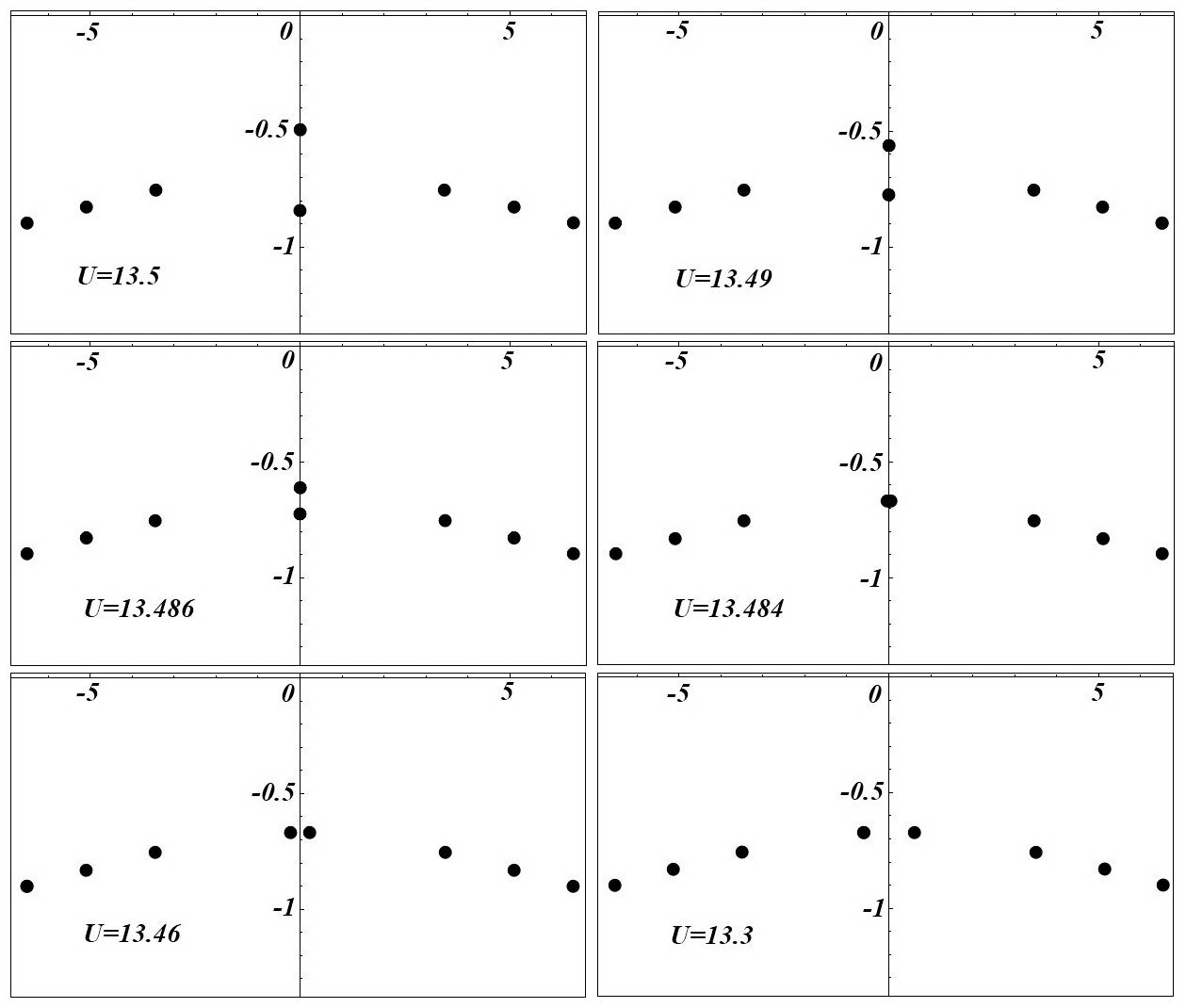}
\caption{\small\textit{\bfseries{Coalescence of $S$-matrix poles. }}\it{Two anti-bound poles
merge into a double pole, that consequently gives rise to a resonance + anti-resonance pair.}}
\label{fig:transition}
\end{center}
\end{figure}
For positive/negative values of $U$ we have a potential well/wall, as shown in  Fig. \ref{fig:potentials}. The matching
conditions for the  wave function and its derivative  are given by \bea\label{matching_general_sym} \ \ \ \ \ \ \ \ \ \ \ \ \ \
\ \ \ \ \ \ \ \ \ \psi(\pm a-0)=\psi(\pm a+0), \  \psi'(\pm a-0)=\psi'(\pm a+0)\ .\ \ \ \ \ \ \ \ \ \ \ \ \ \ \  \ \ \ \eea We
apply (\ref{matching_general_sym}) to the wave function in the well $\psi_U  := A\exp(i K x)+B \exp(- i K x)$ and
(\ref{reflection_WF}), where $K=\sqrt{k^2+2U}$, which allows us to determine the wave function on the whole real line. This way
we also determine the scattering matrix via (\ref{S_matrix}) :
\begin{align}\label{S_matrix_sym}
S=\frac{2e^{-2 i k a}k K}{2 k K \cos(2 K a)-i (k^2+K^2)\sin(2 K a)}
\left(\begin{matrix}
1 &\frac{(k^2-K^2)\sin(2 K a)}{2 i k K}\\
\frac{(k^2-K^2)\sin(2 K a)}{2 i k K}& 1
\end{matrix}\right).
\end{align}
The poles of  the $S$-matrix must obey the nonlinear algebraic equation:
\bea\label{eqn_general_sym}
k\sqrt{k^2+2U}\cos(2 a K)-i(k^2+U)\sin(2 a K)=0.
\eea

They will depend on $a$ and $U$  (and if we consider the asymmetric interval, also on $b$; see Sect. 4 below).  In this section
we consider the symmetric interval $b=-a=1.5$  in dimensionless units.  Fixing the width of the well, we will have a
one-parameter family of solutions of (\ref{eqn_general_sym}). In the complex momentum plane it generates a flow, a collection
of curves representing the positions of the poles as a function of the depth/height $U$, which we will discuss below.

Using Mathematica, Maple or other similar software, it is easy to find all the singularities $k_n$.  We used Mathematica's
built-in functions to find all solutions in a bound region around the origin. Ideally, one can obtain the full flow by
increasing the depth/height incrementally  and generating a pole configuration at each step. The  resulting points can then be
plotted,  displaying the flow. One can plot as many points as desired, making the plot  more detailed or covering wider range
of parameters. We introduce a depth/height cutoff in order for the algorithm to be finite. The cutoff is chosen so that the
plots include the interesting features. At least in theory,  one can decrease the step that changes the parameters to achieve
very high level of detail. In practice, however, the average computer has the capability to generate only a limited number of
configurations. For sufficient detail we plotted small portions of the flow at a time and then combined them.

The singularities of the $S$-matrix are almost always simple poles\footnote{For a discrete set of $U$ values there is a double
pole.} with a typical configuration shown in Fig. \ref{fig:poles}. As expected from the general properties of the scattering
matrix, discussed in the previous section, the singularities  lie on the imaginary momentum axis as well as in the lower
complex plane. Following Nussenzveig \cite{Nussenzveig} we plot the flow (i.e. the  positions of the resonances in the complex
plane) for depth/height  $U \in(-\infty, \infty)$, treating the square potential well/wall, i.e. the well and wall together.

First, let us consider the poles with nonzero real part, i.e. the {\it resonances} (or {\it resonance poles/states}). Recall
that resonances occur in pairs, at two momenta with equal and negative imaginary parts, and equal and opposite real parts.
Those in the third quadrant are known as {\it anti-resonances}. The fourth-quadrant poles are referred to simply as resonances,
when no confusion is likely, and sometimes as ``physical resonances'' when distinctions have to be clear.  The flow is shown in
Fig. \ref{fig:Resonance_symmetric} for the anti-resonances, whose behaviour in time is the time-reversal of that of the
resonances. These suffice for a complete description, because of the symmetry under reflection across the Im$(k)$-axis. When
the well is very deep the poles are located close to the horizontal line ${\rm Im}(k)=-1/a$. As the well becomes shallower the
poles move monotonously downwards along the trajectories as indicated by the arrows in Fig. \ref{fig:Resonance_symmetric}. In
the case of a wall of increasing height,  the poles travel upwards, cross ${\rm Im}(k)=-1/a$ and diverge to the left.

The  horizontal line ${\rm Im}(k)=-1/a$  bounds the pole trajectories for the well. Moreover and as the depth increases a
resonance + anti-resonance pair of poles travels to $k=-i/a$ from each side of the ${\rm Im}(k)$-axis and coalesces to form a
double pole, consequently creating an anti-bound pair,  as in Fig. \ref{fig:transition}. This only happens for  a countable set
of values $U_n, n\in \intg$, that can be found numerically. The point $k=-i/a$  is a coalescence point for the flow and it
represents a transition of a pair of resonance states into a pair of anti-bound states.  For a very deep well the imaginary
part gets closer to the boundary, accounting for the characteristic flattening of the pole distribution for $U\ll 0$.

The discrete subset of values of the depth/height $U_n$, for which a double pole exists, corresponds to each of  the disjoint
sections of the flow: $U_1$ corresponds to the trajectory closest to the Im$(k)$-axis (and its 4th quadrant counterpart), $U_2$
to the second closest, etc. This  is easy to see if we notice that once the resonance reaches the boundary and disappears there
is no other resonance on that trajectory, as the whole resonance configuration just travels up and to the right along fixed
branches of the flow.

Another way of labeling the trajectories is by their asymptotic behaviour. We find from (\ref{eqn_general_sym}) that the pole
locations  $k_n\rightarrow\pm\pi n/2a-i\infty$, $n\in\ntrl$, as $U\to 0$.  The rate at which the poles move also increases as
$U\rightarrow 0$. In fact, as pointed out by Regge \cite{Regge}, very small changes in the depth result in very large shifts in
the locations of the resonances. Despite their sensitivity, however, they follow a smooth trajectory and while the flow rate of
the poles is large for shallower wells, the trajectory is well defined for all values of $U$. As $\vert k_n \vert\rightarrow
\infty$, we verify that resonances are absent in the free particle case.

For large $n$, the locations of the resonance + anti-resonance poles can be approximated \cite{resonances}. For a finite range
constant potential we have ${\rm Re\,} k_n\ \sim\  n$ and ${\rm Im\,} k_n\ \sim\ - {\rm ln}\, n$ as $n\to\infty$. It can also
be shown that the distance between the asymptotic values of the branches of the flow $\lim_{U\rightarrow \infty}(k_n-k_{n-1})$
is independent of $n$.

As discussed, the parameter $U$ can be taken into negative values to reproduce the square wall as a deformation of the square
well.  The flow generated for $U<0$ comes from $-i\infty$ and shares vertical asymptotes with the flow for $U>0$.  For that
reason all the branches are parametrized by their asymptotic values, as before. The shape of the trajectories is shown in Fig.
\ref{fig:Resonance_symmetric}. In the infinite wall limit ($U\rightarrow-\infty$) we find the asymptotic values $\vert
k_n\vert\rightarrow\infty$. Clearly, for  an ``impenetrable'' barrier, resonances do not exist.

Since the flow does not converge to a single point as in the well case, no bound and anti-bound states will be created by the
resonances.  For the attractive potential the resonance states are attracted towards each other to produce bound states. The
location of the attractor (or coalescence point $k=-i/a$) is independent of the potential strength (at least for the cases we
consider) but depends on its width. Conversely, the repulsive potential pushes the resonances away, preventing them from
coalescing, and so from subsequently producing a bound state.

Now let us turn our attention to the singularities with  ${\rm Re}(k)=0$.  They correspond to the bound states $({\rm
Im}(k)>0)$ and the anti-bound states $({\rm Im}(k)<0)$. For a potential well, consider a typical bound-state--not the ground
state, nor the first excited state. The corresponding pole location has $({\rm Im}(k)>0)$, and as the well-depth decreases,
$({\rm Im}(k)$ decreases monotonically. When $({\rm Im}(k)$ becomes negative, the state changes from bound to anti-bound. For a
small range of $U$ thereafter, it has the strange property of an ``energy'' that becomes more negative as the well becomes
shallower. Then it meets another anti-bound state ascending from $-i\infty$ at the $k=-i/a$ coalescence point. There a pair of
anti-bound states ``annihilate'' giving birth to a resonance + anti-resonance pair of poles (Fig. \ref{fig:transition}). The
decrease in the number of bound states is tied to the decrease in the number of anti-bound states. The exceptions are  the last
two bound-state poles--one having $0$ as a limit and the other rapidly traveling to $-i\infty$. For very shallow wells, a
surviving bound state always exists. The last bound state becomes an anti-bound when the well turns into a wall, and heads to
$- i\infty$ as the wall grows, with the Im$(k)$ axis as an asymptote.

One can see that as the well becomes shallower the highest bound state will turn into a zero-energy state.  Zero-energy states
have infinite spread, i.e. their de Broglie wave length becomes infinite. They represent a critical point: the normalizable
wave functions become divergent anti-bound states just above the well. From (\ref{eqn_general_sym}), we see that zero-energy
states are admitted when $U$ takes the values ${n^2 \pi^2}/{8a^2},\ n\in \N$.

Besides the bound and anti-bound state flows,  Fig. \ref{fig:symmetric_bound} also depicts the trajectories of the branch
points, distinguished by the hollow points. The  branch cut is determined by $\sqrt{k^2+2U}$, i.e.
Im$(k)\in(-\infty,-\sqrt{2U})\cup( \sqrt{2U},\infty)$. As we can see all the bound states are contained within the parabola,
their energy cannot be smaller than the depth of the well, since the wave functions for  $E<U_{min}$ are not normalizable. As
the anti-bound states are non-normalizable anyway, they are not affected by the branch cut, as  the anti-bound states for
$-1<U<0$ illustrate in Fig. \ref{fig:symmetric_bound}. An interesting observation is that the bound/anti-bound pairs are always
contained within the branch points for all values of $U$.

For the finite barrier, i.e. $U<0$,  only resonance (and anti-resonance) poles occur. Recall that when the resonance flow
converges, a pair of trajectories collide from each side of the imaginary axis which results in the creation of a
bound/anti-bound pair. For the potential wall   the resonance flow diverges, so there are no bound states, and therefore, no
anti-bound states either.


\section{Flows for the two-piece well+wall  potential}
\begin{figure}[t]
\begin{center}
 \includegraphics[width=120mm]{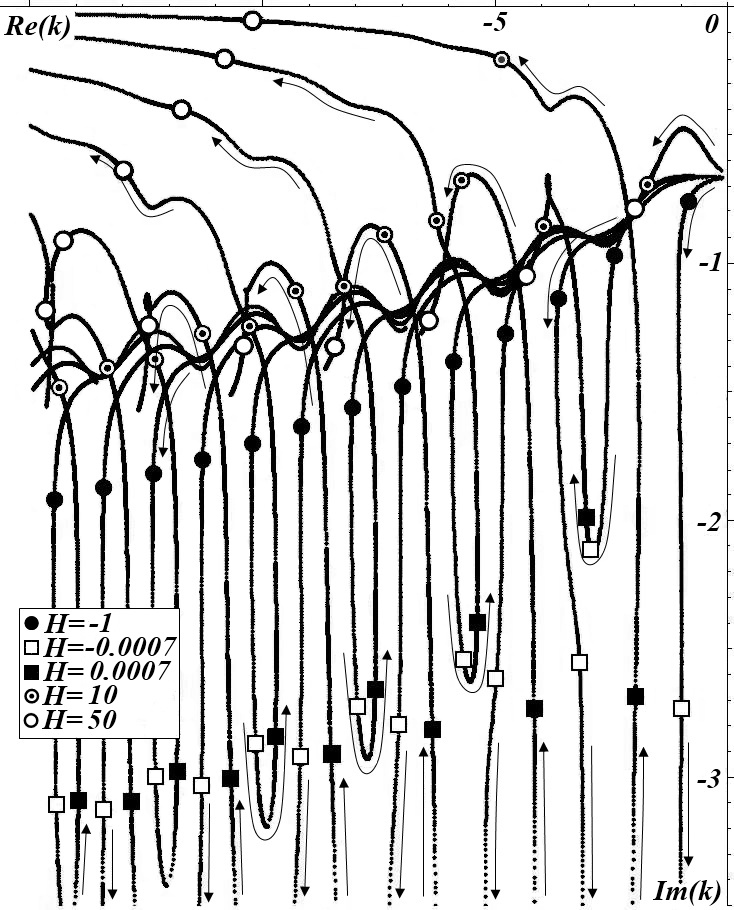}
\caption{\small\textit{\bfseries{Flow trajectories of the  anti-resonance states. }}\it{The flow is for a potential
consisting of a fixed well and variable well changing into a wall (fixed $U$, varying $H$).}}
\label{fig:asymmetric_resonances}
\end{center}
\end{figure}
\begin{figure}[t]
\begin{center}
 \includegraphics[width=120mm]{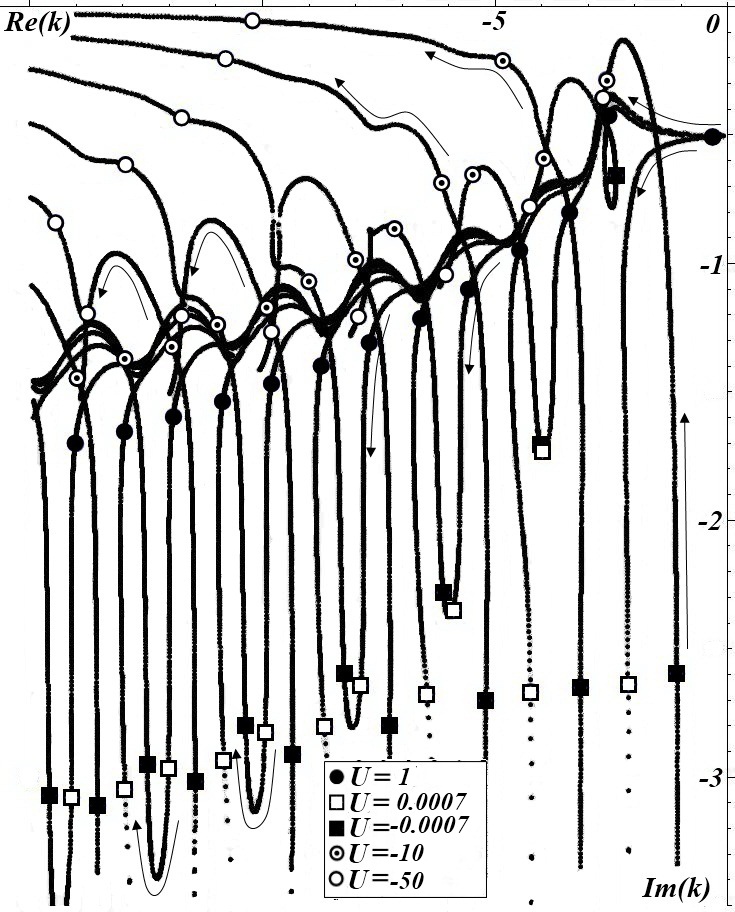}
\caption{\small\textit{\bfseries Flow trajectories of the  anti-resonance states.  }\it{The flow is for a potential consisting
of a fixed wall and variable well changing into a wall (fixed $H$, varying $U$).}}
\label{fig:reversed_res}
\end{center}
\end{figure}
Now let us investigate how the presence of both attractive and repulsive parts affects the flow of the poles of the $S$-matrix.
Again, we  consider the simplest case possible, a potential that has two constant non-zero pieces, the well+wall  shown in Fig.
\ref{fig:potentials}. \bea\label{asymmetric_well} V(x)=\Biggl\lbrace
\begin{matrix} &-U, \ x \in[a,0],\qq\ \ \ \ \ \ \,\, \cr
 &H, \ x \in(0,b],\qq\ \ \ \ \ \, \cr
&0,\  x\in(-\infty,a) \cup (b,\infty).\, \cr
\end{matrix}
\eea When either $U$ or $H$ change sign, the corresponding piece of the potential changes character, from a well to a wall, or
vice-versa. Therefore, to illustrate the changes to the pole distribution, we will provide 2 graphs: Fig. 6 depicts the flow
for the well+well/wall, and Fig. 7 the wall+well/wall.

The matching at the discontinuities provides the analytical expression for the $S$-matrix. Using the ansatz
(\ref{reflection_WF}) we find  the scattering matrix in the form of (\ref{S_matrix}):
\begin{align}\notag
&T=2 e^{i (a - b) k} k \kappa K/{\cal T} \\ \label{asymm_S_matrix}
&R_+=-e^{2 i a k}\bigl[ \kappa (K^2-k^2)\sin(a K)\cos(b \kappa)+\\ \notag
&\QQ +i k (\kappa^2-K^2)\sin(a K)\sin(b \kappa)+K (K^2-\kappa^2)\cos(a K)\sin(b \kappa)\bigr]/{\cal T} \\ \notag
&R_-=-i e^{-2 i b k}\bigl[ \kappa (k^2-K^2)\sin(a K)\cos(b \kappa)-\\ \notag
&\QQ -i k (\kappa^2-K^2)\sin(a K)\sin(b \kappa)+K (\kappa^2-k^2)\cos(a K)\sin(b \kappa)\bigr]/{\cal T}\notag
\end{align}
with $\kappa = \sqrt{k^2 - 2 H}, K = \sqrt{k^2+2U}$ and
\bea\label{poles_equation_asymm}
{{\cal T} }=\sin(b \kappa) \bigl[-i (\kappa^2 + k^2) K \cos(a K) + K (\kappa^2 + K^2) \sin(a K)\bigr] +\ \ \ \ \ \ \ \ \ \ \ \ \ \ \  \ \ \\ \notag
\kappa \cos(b \kappa) \bigl[2 k K \cos(a K) +i(k^2 + K^2) \sin(a K)\bigr].\notag
\eea
\begin{figure}
\begin{center}
 \includegraphics[width=140mm]{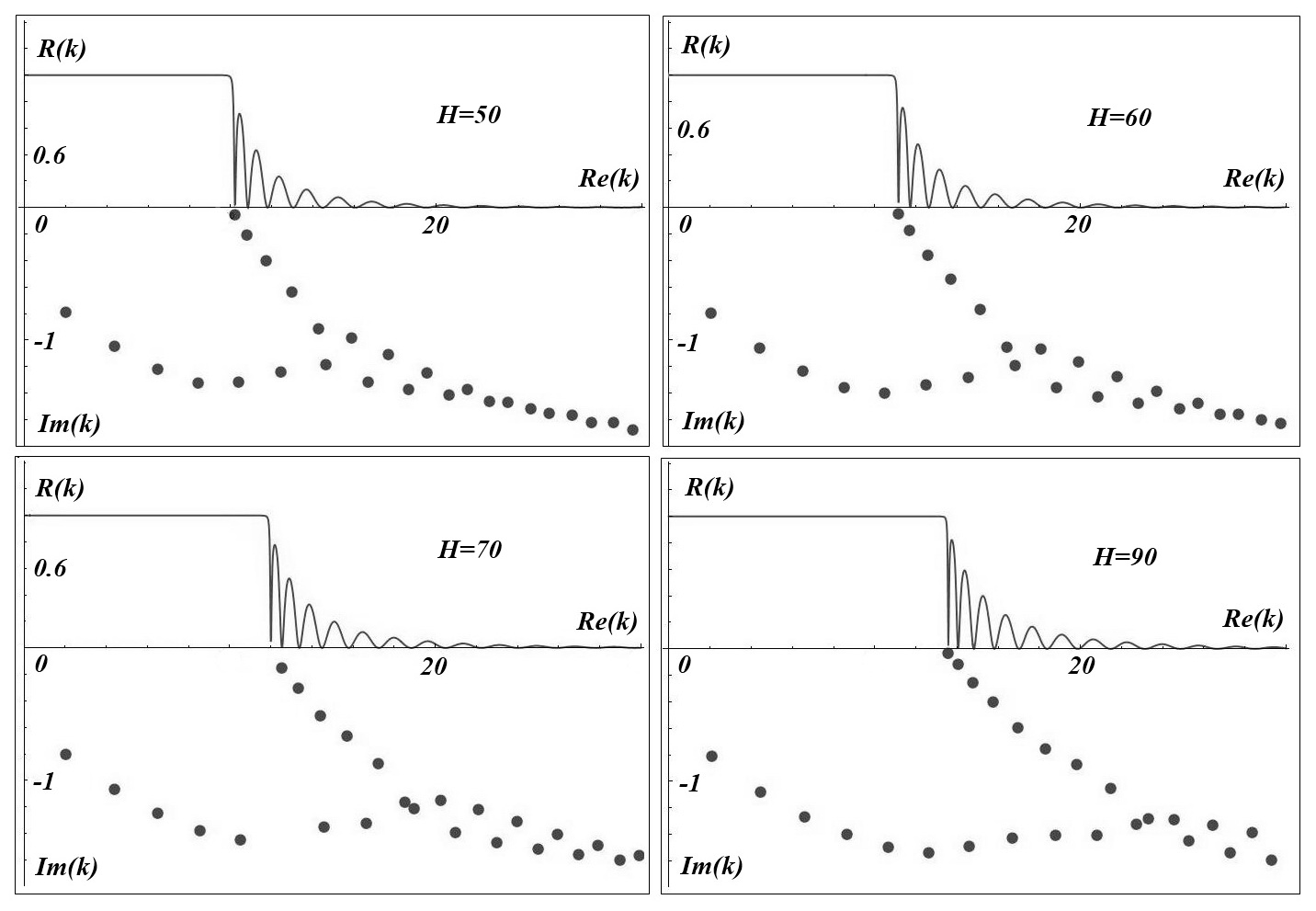}
\caption{\small\textit{\bfseries Resonance poles and the reflection coefficient. }
\it{The well+wall potential has a well with fixed depth with the barrier  increasing in height. Increasing the wall
height ``unzips'' the line of resonances into two lines. }}
\label{fig:large_potential}
\end{center}
\end{figure}
\begin{figure}
 \begin{center}
\includegraphics[width=140mm]{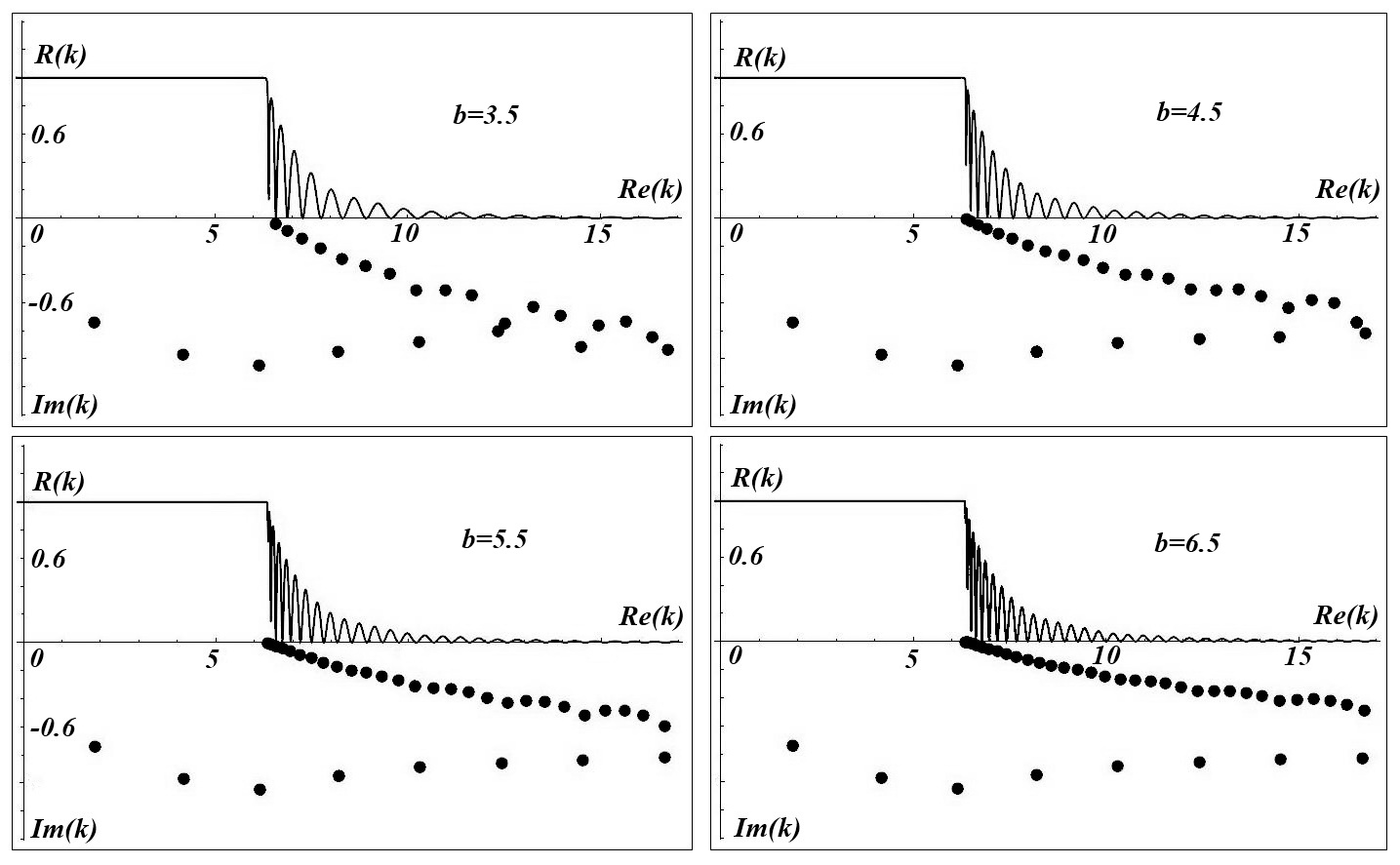}
\caption{\small\textit{\bfseries Resonance poles and the reflection coefficient. }
\it{The well+wall potential has a well with fixed depth  and a wall with fixed height, but increasing width.
The ``unzipping'' also occurs in this case. }}
\label{fig:wide_potential}
\end{center}
\end{figure}
\begin{figure}
\begin{center}
 \includegraphics[width=95mm]{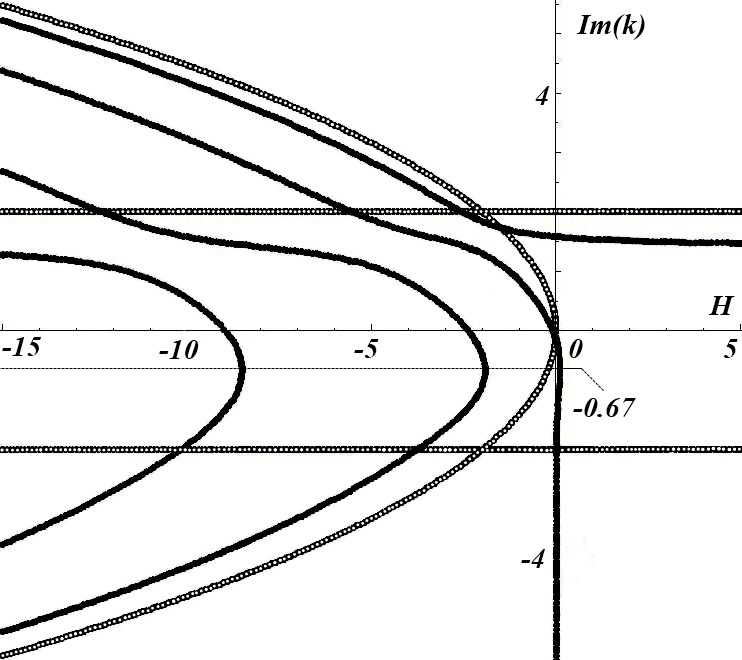}
\caption{\small\textit{\bfseries Flow of the bound and anti-bound poles for the well+wall potential.}
\it{The potential consists of a well with fixed depth and a well transforming into a wall with
variable depth/height (fixed well + variable well/wall).}}
\label{fig:asymmetric_bound}
\end{center}
\end{figure}
\begin{figure}
\begin{center}
 \includegraphics[width=90mm]{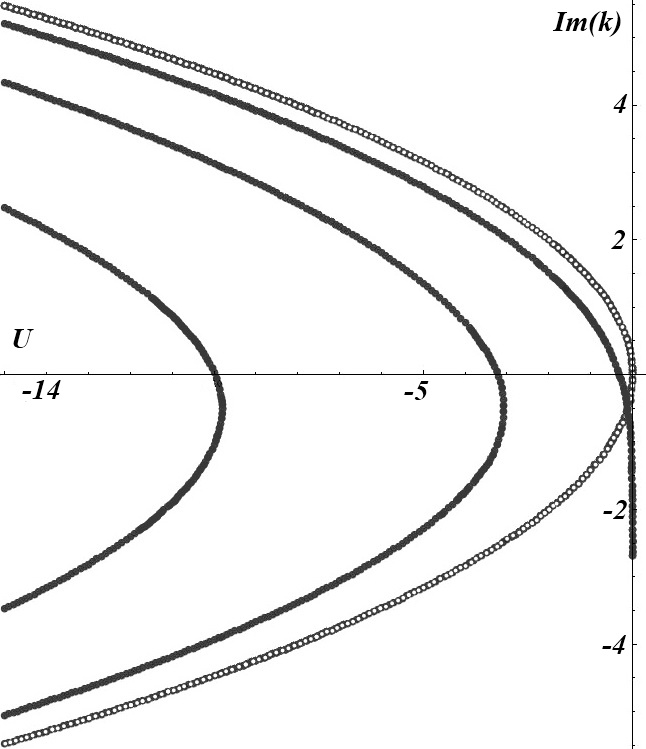}
\caption{\small\textit{\bfseries  Flow of the bound and anti-bound poles for the well+wall potential.}
\it{The potential consists of a wall with fixed height and a well transforming into a wall with
variable depth/height (fixed wall + variable well/wall).}}
\label{fig:reversed_bound}
\end{center}
\end{figure}
\begin{figure}
\begin{center}
\includegraphics[width=90mm]{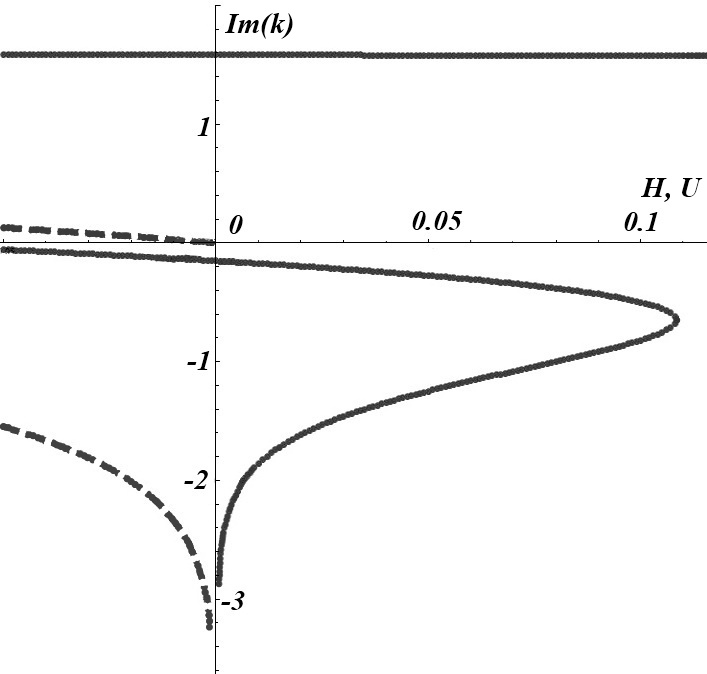}
\caption{\small\textit{\bfseries Magnified flow for the 2-piece potential. }
\it{Trajectories of the bound,  and anti-bound states for the fixed well part (dashed line) and
fixed wall (solid line)  potentials around the origin.}}
\label{fig:exaggerated}
\end{center}
\end{figure}

The poles are then determined by the equation ${{\cal T} }=0$. The resonance flows are shown in Figs.
\ref{fig:asymmetric_resonances} and \ref{fig:reversed_res}. In the first case we have a potential that consists of a fixed well
of depth $U$ combined with a  variable-depth ($H<0$)  well transitioning into a wall ($H>0$). The second figure (Fig.
\ref{fig:reversed_res}) shows the ``reverse'' case, where we fix the barrier and vary the well with depth $-U<0$ into a barrier
with height $-U>0$.

Let us concentrate on the first case for now. The flows are now separated into two distinct families. For a fixed well + varied
well we have the familiar behavior for very deep (symmetric) well--all the poles tend to move towards $k=-i/a$, with the
characteristic annihilation of resonances occurring as before. However the presence of the fixed well destroys the monotonous
behavior of the flow and generates the oscillating pattern seen in both Fig. \ref{fig:asymmetric_resonances} and Fig.
\ref{fig:reversed_res}. The oscillating pattern shows that despite the fact that $\vert U\vert\ll\vert H\vert$, there is an
observable influence of the finite part even when the variable part of the potential is of much greater magnitude.

$H\rightarrow0$ shows separation of the resonances into two families: rapidly diverging, due to the increasing wall, and slowly
moving (finite well) ones. As we will confirm later, each family can be associated with one of the two pieces of the potential.
The slowly moving half of the resonances need to recover the configuration for a finite well (of $1/2$ the width) so they
remain finite for all $H$. The transition from $H<0$  to $H>0$ result in the local minima of Im$(k)$ observed in  Fig.
\ref{fig:asymmetric_resonances}.

As the well turns into a wall we observe rapid growth of new branches from $-i\infty$. Those consist of poles that can be
associated with the varied part of the potential. Unlike the symmetric well, however, the flow does not share asymptote with
its $H<0$ counterpart.\footnote{We have to truncate the values of $H$, in order for the algorithm to terminate, but we did not
observe convergence between the $H<0$ and $H>0$ branches of the flow.}

Again we observe two types of resonance poles. On one hand we observe that some of the poles follow  diverging trajectories
heading  to the left with Re$(k)\rightarrow-\infty$ and Im$(k)\rightarrow0$. The locations of the rest of the poles move
upwards until they reach maxima and then descend with decreasing rate to reach a configuration that changes very slightly with
the varied height of the barrier. However we observed similar behavior for the 1-piece potential barrier. This seems to suggest
that indeed the poles $\lim_{H\rightarrow\infty} \vert k\vert \rightarrow  \infty$ correspond to the barrier and the rest -- to
the well. Further confirmation can be found in  Fig. \ref{fig:large_potential} -- it shows ``snap shots'' of the locations of
the poles. The pole configuration exhibits a zipper-like behavior due to the separation of the poles along the two distinct
flow branches, as discussed. The ``top'' poles are pushed away by the branch cut of $\sqrt{k^2 - 2 H}$ (introduced by the
varied well/barrier)  and  does not affect  the ``bottom'' poles. In addition Fig. \ref{fig:wide_potential} shows that widening
the barrier, for fixed $H$ does not affect the ``bottom'' poles while significantly increases the count of the ``top'' poles.

The pole separation is quite surprising since the behaviour of the resonances is determined by a non-linear relation and yet
they behave in a remarkably linear fashion. Only when the  two pieces are of comparable size can the resonances not be
distinguished without information about their trajectories. However, labeling the poles will be incorrect since the poles
corresponding to the fixed well and those corresponding to the varied well barrier can exchange roles and only make sense when
$\vert U\vert\ll\vert H\vert$. Indeed the poles that survive the limit $H\rightarrow 0$ are on the same trajectory as the poles
that vanish in the limit $H\rightarrow \infty$.

Note that Fig. \ref{fig:wide_potential} indicates, as mentioned earlier, that the ``top''  resonances correspond to local
minima of the reflection probability. Thus, the name resonance poles is justified--they are fully transmitted modes responsible
for a transmission resonance. For low potential barriers the peaks begin to widen and overlap, as in the case of the ``bottom''
poles, gradually destroying the visual correspondence observed in Fig. \ref{fig:wide_potential}. The last two figures also
illustrate the physical significance of the resonance poles, as well as the branch cuts. The resonances indicate maxima in the
transmission coefficients, while the brunch cuts act as an impenetrable barrier for the resonance states associated with the
varied portion of the potential.

Now let us consider the bound states. Fig. \ref{fig:asymmetric_bound} shows that when two wells interact the matching distorts
the trajectories and they have inflexion points for certain values of the potential. This ``complex plane'' scattering is not
present in the case of Fig. \ref{fig:reversed_bound} since the barrier does not have bound states. While there are also
zero-energy states for a discrete set of potentials, they cannot be found in closed terms, due to the more complicated matching
conditions.

There is a noticeable difference around $U=0$--instead of the least-bound state turning changing into an anti-bound one and
then disappearing to infinity, we have an ascending anti-bound state annihilating it for a very small height of the wall. The
most-bound state  survives the transition from a well to a wall due to the presence of the fixed well to the left.

The features around the zero potentials are interesting enough to be plotted in greater detail in Fig. \ref{fig:exaggerated}.
Apparently, the absence of bound states associated with the barrier part leaves the familiar monotonous  behavior of the
bound/anti-bound states unchanged.

 \section{Conclusion}

We studied the flow of $S$-matrix poles in the plane of complex momentum for the simple potential(s) depicted in Fig.
\ref{fig:potentials}.  Our results are summarized by the remaining Figures.

The most elementary potential considered was the square well/wall, and our Figs. \ref{fig:poles}-\ref{fig:transition} recover
the seminal results of Nussenzveig \cite{Nussenzveig}.  Some of Nussenzveig's conclusions were also confirmed by less direct,
graphical methods in \cite{Moiseyev} (see also \cite{Kawasaki}).

New is a similar study of the ``two-piece'' potential, consisting of two adjacent well/walls, with strengths controlled by
independent parameters ($U$ and $H$ in Fig. \ref{fig:potentials}).  The flow of $S$-matrix poles was calculated as a function
of various potential parameters, one at a time, and plotted in Figs. \ref{fig:asymmetric_resonances}-\ref{fig:exaggerated}.

\begin{figure}
\begin{center}
 \includegraphics[width=120mm]{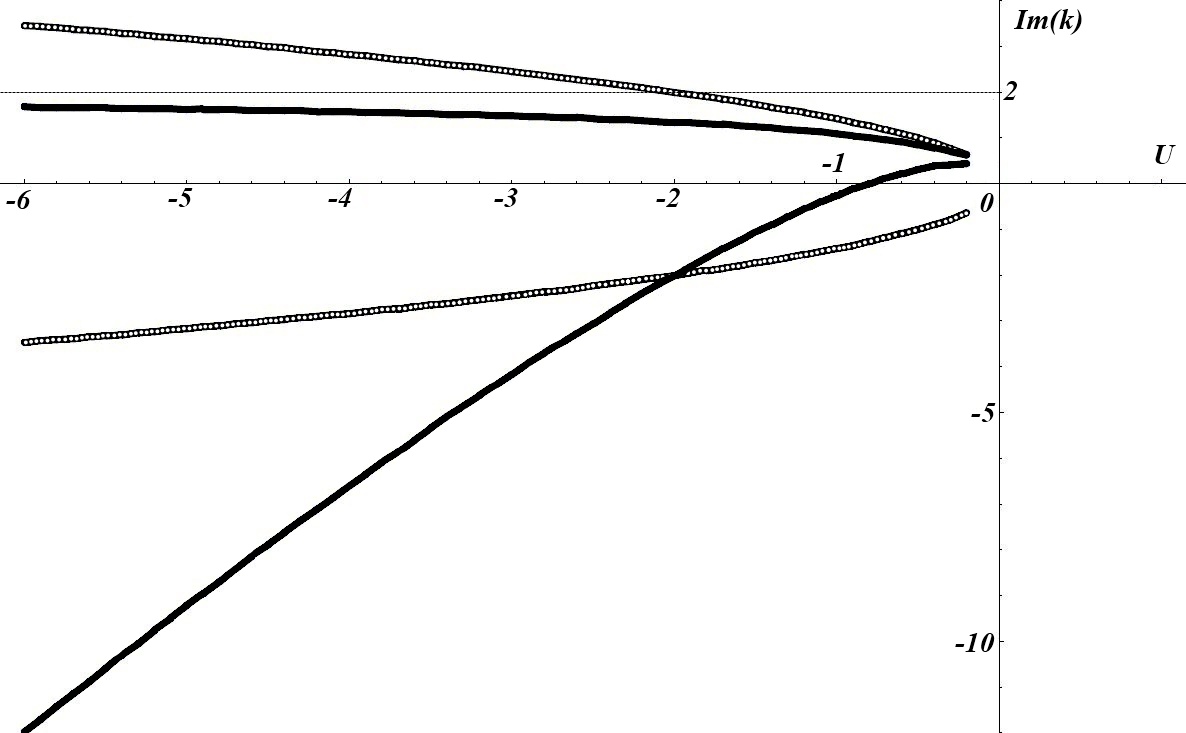}
\caption{\small\textit{\bfseries Flow of the  bound/anti-bound states of the  $\delta$-sequence  potential. }
\it{A single bound state survives as the potential approaches the delta-function limit.}}
\label{fig:delta_bound}
\end{center}
\end{figure}

\begin{figure}
\begin{center}
 \includegraphics[width=120mm]{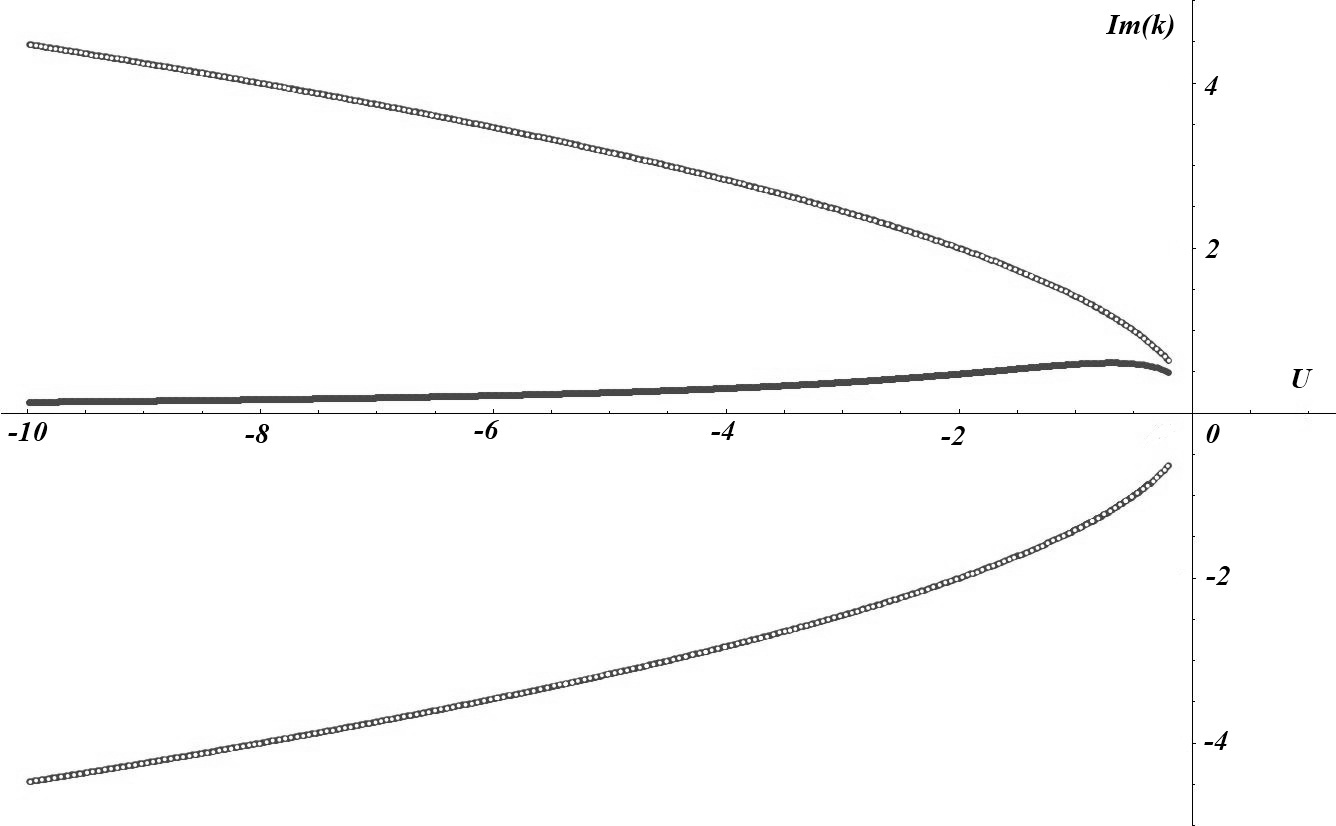}
\caption{\small\textit{\bfseries Flow of the  bound state of the  $\delta'$-sequence  potential. }
\it{A single bound state survives; however, it is a zero-energy state.}}
\label{fig:delta_prime_bound}
\end{center}
\end{figure}

\begin{figure}[t]
\begin{center}
 \includegraphics[width=\linewidth]{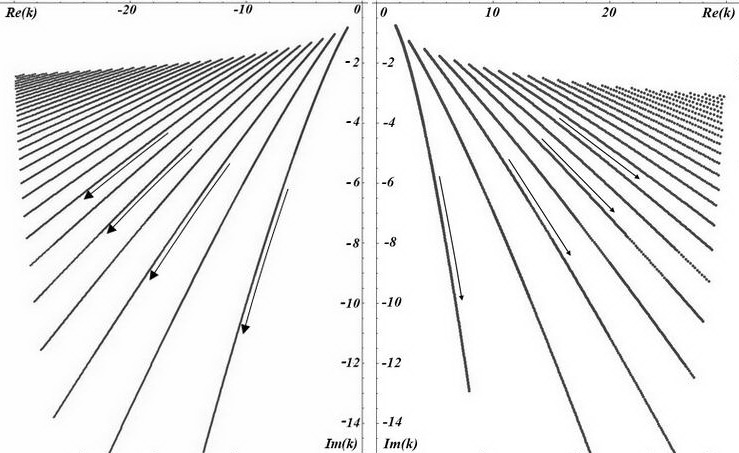}
\caption{\small\textit{\bfseries Flow of the  resonance states in the  $\delta$-sequence (left) and $\delta'$-sequence (right) potentials. }
\it{All the resonances move away from the origin, completely disappearing in the two limits.}}
\label{fig:delta_delta_prime_together}
\end{center}
\end{figure}

The Appendix also verifies that the ``one-piece'' and ``two-piece'' flows are consistent with the pole structure of a
$\delta$-function potential and a $\delta'$-function potential, respectively, if the  parameters are controlled in the
appropriate way. The calculations are a simple, yet interesting check of our methods.   Figs.
\ref{fig:delta_bound}-\ref{fig:delta_delta_prime_together} depict the relevant flows.

Unsurprisingly, the flows for the more complicated potentials are more complicated.  Certain features can be understood simply,
however, and we hope that our extension of the old analysis of Nussenzveig \cite{Nussenzveig} will help lead to further insight
that can be applied generally.


\vskip2.5cm {\bf\large{\noindent Appendix:\ {$\delta$- and $\delta'$-sequence potentials}}}

We can perform a simple, interesting check of our results by recovering the well-known textbook examples of the Dirac delta
function (or $\delta$) potential and its derivative, $\delta'$, as limits of the ``one-piece'' and ``two-piece'' potentials,
respectively.

Consider first $V(x)=-\lambda \delta(x)$. We realize this $\delta$-potential as the $a\to 0$ limit of a $\delta$-sequence of
the ``one-piece" potential:  $-U\,\left[\theta(x+a)-\theta(x-a)\right]$, with the strength $\lambda= 2 a U$ kept fixed. To see
what we should find, we solve the Schr\"odinger equation for $x\in\real/\{0\}$ and demand that the wave function is
square-integrable to show $\psi\propto e^{-\lambda\vert x\vert}$. Integrating equation (\ref{eigenvalue}) over an infinitesimal
interval $(-\varepsilon, \varepsilon)$ around $0$ yields \ben \psi'(\varepsilon)-\psi'(-\varepsilon)=-2m\lambda\psi(0),
\label{delta_bcs} \een which determines the energy $E=-m \lambda^2/2$ of the bound state. A straightforward calculation shows
the transmission and reflection  amplitudes \ben t=(1-m\lambda/ik)^{-1},\ r=(ik/m\lambda-1)^{-1}\label{trans_refl} \een have a
pole only at the bound state energy, i.e. no resonances exist for the   $\delta$-potential. This is also confirmed by the fact
that the probabilities for  transmission $\vert t \vert^2$ and reflection $\vert r \vert^2$ are monotonous functions, i.e.
there are no transmission maxima.

The graphical representation of the flow leads to the same conclusions.  The pole spectrum of the square well flows
asymptotically to that of the $\delta$-potential. Setting $a U=1$ and taking the limit $U\rightarrow\infty$ produces the
trajectories of the resonance poles shown on the left side of Fig. \ref{fig:delta_delta_prime_together}.  The resonances
approach complex infinity as the depth increases and no finite resonances exist in the $\delta$-limit. Similarly, a single
bound state survives (Fig. \ref{fig:delta_bound}) in agreement with the above discussion. The final anti-bound  state rapidly
diverges to $-i\infty$ leaving no trace.

Now let us consider the derivative of the $\delta$-potential as arising from \ben -U\left[\theta(x+a) - 2\theta(x) +
\theta(x-a)\right]\ ,\een a special case of the ``two-piece potential''. We can show that  the Schr\"odinger equation with
$\delta'$-potential has the same solutions as for the $\delta$-potential  but the matching has to be done differently. Assuming
the wave function is continuous at $x=0$, we once again integrate the Schr\"odinger equation around zero in order to obtain a
matching condition for the derivative: \ben \psi'(-\varepsilon)=\psi'(\varepsilon). \een This is quite different from the
$\delta$-potential, since the matching condition demands a wave function that is also smooth at the origin, which is only
possible for $k=0$. Matching for a plane wave incident from the left and right shows the absence of resonances and
anti-resonances as well.

While somewhat trivial, this potential illustrates another zero-feature  limit, that of the well+wall. The flows generated by
the limit are shown in Fig. \ref{fig:delta_prime_bound} for the bound/anti-bound poles and  Fig.
\ref{fig:delta_delta_prime_together} (together with the $\delta$ case), in agreement with the analytic arguments.

\end{document}